\documentstyle[epsf,preprint,tighten,prd,aps]{revtex}

\newcommand{\Lie}{{\pounds}}
\renewcommand{\Re}{\text{Re} }
\renewcommand{\Im}{\text{Im} }

\renewcommand{\L}{{\cal L}}
\newcommand{\D}{{\cal D}}
\newcommand{\F}{{\cal F}}

\begin{document}
\epsfverbosetrue
\draft


\title{
  Continuous Self-Similarity Breaking in Critical Collapse
}
\author{Andrei V. Frolov%
  \thanks{Email: \texttt{andrei@phys.ualberta.ca}}}
\address{
  Physics Department, University of Alberta\\
  Edmonton, Alberta, Canada, T6G 2J1}
\date{16 August, 1999}
\maketitle

\begin{abstract}
  This paper studies near-critical evolution of the spherically
  symmetric scalar field configurations close to the continuously
  self-similar solution. Using analytic perturbative methods, it is
  shown that a generic growing perturbation departs from the Roberts
  solution in a universal way. We argue that in the course of its
  evolution, initial continuous self-similarity of the background is
  broken into discrete self-similarity with echoing period $\Delta =
  \sqrt{2}\pi = 4.44$, reproducing the symmetries of the critical
  Choptuik solution.
\end{abstract}

\narrowtext


\section{Introduction} \label{sec:intro}

Critical phenomena in gravitational collapse have been a relatively
recent and interesting development in the established field of general
relativity. Following the numerical work of Choptuik on the spherically
symmetric collapse of the minimally coupled massless scalar field
\cite{Choptuik:93}, critical behavior was discovered in most common
matter models encountered in general relativity, including pure gravity
\cite{Abrahams&Evans:93}, null fluid \cite{Evans&Coleman:94} and, more
generally, perfect fluid \cite{Koike&Hara&Adachi:95,Maison:96}, as well
as more exotic models. \nocite{Hara&Koike&Adachi:96}

The essence of critical phenomena in general relativity is the fact
that just at the threshold of black hole formation, the dynamics of the
field evolution becomes relatively simple and, in some important
aspects, universal, despite the complicated and highly non-linear form
of the equations of motion. In analogy with second order transitions in
condensed matter physics, the mass of the black hole produced in
near-critical gravitational collapse scales as a power
law\footnote{Usually, but there are models with mass gap in black hole
production, most notably Yang-Mills field, whose behavior is more
analogous to a first order phase transition.}
\begin{equation} \label{eq:scaling}
  M_{\text{BH}}(p) \propto |p-p^*|^\beta,
\end{equation}
with parameter $p$ describing initial data, and the mass-scaling
exponent $\beta$ is dependent only on the matter model, but not on the
initial data family. The critical solution, separating solutions with
black hole formed in the collapse from the ones without a black hole,
also depends on the matter model only, and serves as an intermediate
attractor in the phase space of solutions. It often has an additional
symmetry called self-similarity, in either continuous or discrete
flavors.

Discovery of critical phenomena in gravitational collapse was the first
real success of numerical relativity in which a physical effect was
observed in simulations without being first predicted by theoretical
physicists. For the theoretician, the challenge and attraction of
studying critical phenomena lies in the possibility of exploring a new
class of exact solutions of Einstein's equations, having simple
properties and high symmetry, but previously undiscussed. Another
interesting thing about critical solutions is that they are very
relevant to the cosmic censorship conjecture, the long-unsolved problem
of general relativity. With their ability to produce arbitrarily small
black holes and, in the critical limit, curvature singularity without
an event horizon, in the course of quite generic gravitational
collapse, they may serve as an acceptable counterexample to the cosmic
censorship conjecture (see \cite{Gundlach:97:review} and references
therein).

Universality of the near-critical behavior has been explained by
perturbation analysis and renormalization group ideas 
\cite{Evans&Coleman:94,Koike&Hara&Adachi:95,Maison:96,Hara&Koike&Adachi:96},
and is rooted in the fact that critical solutions generally have only
one unstable perturbation mode. In the course of evolution of the
near-critical initial field configuration, all the perturbations modes
contained in it decay, forgetting details of the initial data and
bringing the solution closer to critical, except the single growing
mode which will eventually drive the solution to black hole formation
or dispersal. In this sense, the critical solution acts as an
intermediate attractor in the phase space of all field configurations.
Because there is only single growing mode, the codimension of the
attractor is one. The eigenvalue of the growing mode determines how
rapidly the solutions will eventually depart from critical, and it can
be used to calculate the mass-scaling exponent $\beta$.

As we have mentioned, the critical solution often has additional
symmetry besides the usual spherical symmetry, called continuous or
discrete self-similarity. This symmetry essentially amounts to the
solution being independent of (in case of continuous self-similarity)
or periodic in (in case of discrete self-similarity) one of the
coordinates, a scale. The role of this symmetry in critical collapse is
not understood at all. Some attempts at finding critical solutions made
a continuously self-similar ansatz and hit a jackpot
\cite{Evans&Coleman:94}, whereas others studied discretely self-similar
solutions from phenomenological point of view \cite{Gundlach:96}. Yet
the simple question of why a particular matter model should have this
or that version of self-similarity incorporated in the critical
solution still remains a mystery.

This paper attempts to shed some light on the subject by investigating
the dynamics of formation of discretely self-similar structure in the
gravitational collapse of a minimally coupled massless scalar field. As
a base point of our investigation, we consider a certain continuously
self-similar solution, known as the Roberts solution, as a toy-model of
the critical solution in the gravitational collapse of a scalar field.
This solution was constructed as a counterexample to the cosmic
censorship conjecture \cite{Roberts:89} and was later rediscovered in
the context of critical gravitational collapse
\cite{Brady:94,Oshiro&Nakamura&Tomimatsu:94}. While not a proper
attractor \cite{Frolov:97}, this simple solution resembles in some of
its properties more complicated critical solutions known only
numerically. The aim of the present work is to show how, at least in
the linear approximation, the discretely self-similar structure arises
dynamically in the scalar field collapse. The advantage of our approach
is that, due to the simple form of the Roberts solution, calculations
can be carried out analytically and so provide additional independent
insights different from numerical treatments.

Using linear perturbation analysis and Green's function techniques, we
study evolution of the spherically symmetric scalar field
configurations close to the continuously self-similar solution.
Approximating late-time evolution via the method of stationary phase,
we find that a generic growing perturbation departs from the Roberts
solution in a universal way. In the course of the evolution, initial
continuous self-similarity of the background is broken into discrete
self-similarity by the growing perturbation mode, reproducing the
symmetries of the Choptuik solution. We are able to calculate the
echoing period of the formed discretely self-similar structure
analytically, and its value is close to the result of numerical
simulations.

\section{The Roberts Solution} \label{sec:roberts}

\begin{figure}
  \begin{center}\begin{tabular}{cc}
    \epsfxsize=\textwidth \multiply\epsfxsize 4 \divide\epsfxsize 10
    \epsfbox{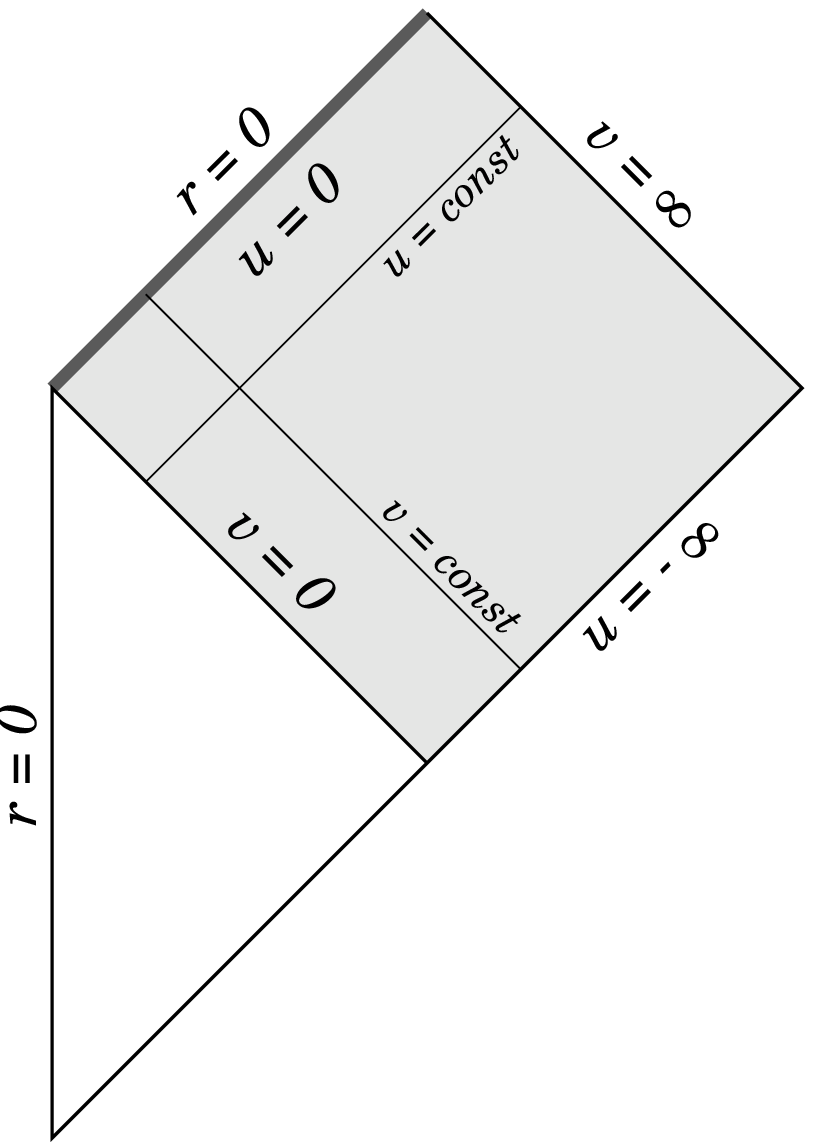} &
    \epsfxsize=\textwidth \multiply\epsfxsize 4 \divide\epsfxsize 10
    \epsfbox{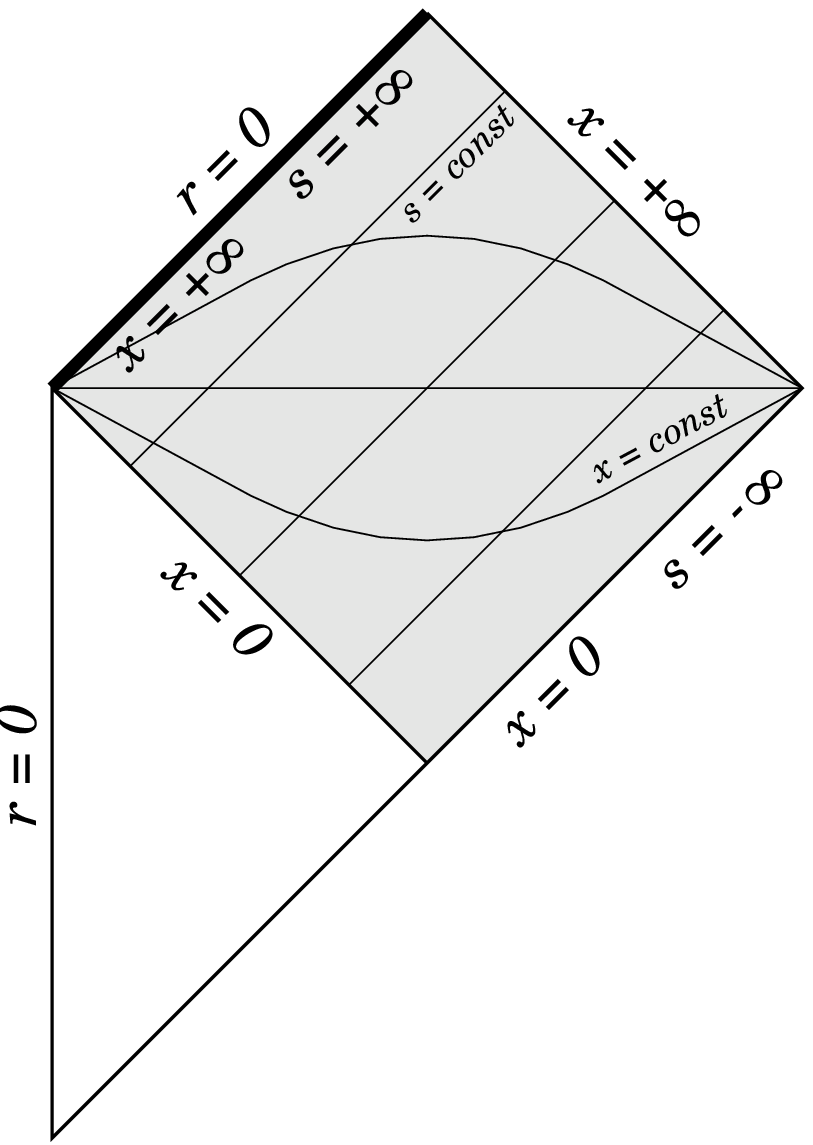}\\
    (null coordinates) & (scaling coordinates)
  \end{tabular}\end{center}
  \caption{
    Global structure of the Roberts solution: The scalar field influx
    is turned on at $v=0$; spacetime is flat before that. The field
    evolution occurs in the shaded region of the diagram, and there is
    a null singularity in the center of the spacetime.
  }
  \label{fig:roberts}
\end{figure}

The starting point of our investigation is the Roberts solution, which
will serve as a background for linear perturbation analysis. It is a
solution describing gravitational collapse of a minimally coupled
massless scalar field, described by the Einstein-scalar field equations
\begin{eqnarray}
  R_{\mu\nu} &=& 2 \phi_{,\mu} \phi_{,\nu}, \label{eq:r}\\
  \Box\phi &=& 0 \label{eq:box},
\end{eqnarray}
which is spherically symmetric and also continuously self-similar. The
latter symmetry means that there exists a vector field $\xi$ such that
\begin{equation} \label{eq:ss}
  \Lie_\xi g_{\mu\nu} = 2g_{\mu\nu}, \hspace{1em}
  \Lie_\xi \phi = 0,
\end{equation}
where $\Lie$ denotes the Lie derivative. Under these assumptions the
field equations can be solved analytically, which is most easily done
in null coordinates \cite{Brady:94,Oshiro&Nakamura&Tomimatsu:94,Frolov:98}.
Self-similar solutions form a one-parameter family. As the parameter is
varied, spacetimes both with and without a black hole occur. The
solution just at the threshold of black hole formation\footnote{In the
early works \cite{Brady:94,Oshiro&Nakamura&Tomimatsu:94} term {\it
critical} has been used to designate this solution. This is somewhat
confusing because, strictly speaking, this solution is not an
intermediate attractor of codimension one, and so is not critical in
the usual sense. Perhaps the term {\it threshold} gives better
description of its nature. In any case, since we are not concerned with
the other solutions from the self-similar family in this paper, we will
refer to the solution (\ref{eq:metric},\ref{eq:crit}) by the name ``the
Roberts solution''.} is given by the metric
\begin{equation} \label{eq:metric}
  ds^2 = - 2\, du\, dv + r^2\, d\Omega^2,
\end{equation}
where
\begin{equation} \label{eq:crit}
   r = \sqrt{u^2 - uv}, \hspace{1em}
  \phi = \frac{1}{2} \ln \left[1 - \frac{v}{u}\right].
\end{equation}
The global structure of the corresponding spacetime is shown in
Fig.~\ref{fig:roberts}. The influx of the scalar field is turned on at
the advanced time $v=0$, so that the Roberts spacetime is smoothly
matched to Minkowskian flat spacetime to the past of this surface. The
junction conditions, required for continuity of the solution there,
serve as boundary conditions for the field equations (\ref{eq:r}) and
(\ref{eq:box}). More detailed discussion of this important point is
provided in Appendix~\ref{sec:junct}.

The evolution of perturbations of the Roberts solution is most easily
followed in a coordinate system exploiting scale-invariance of the
background, so that the self-similarity becomes apparent. Therefore, we
introduce new coordinates, which we will call scaling coordinates, by
\begin{equation} \label{eq:coord:xs}
  x = \frac{1}{2} \ln \left[1 - \frac{v}{u}\right], \hspace{1em}
  s = - \ln(-u),
\end{equation}
with the inverse transformation
\begin{equation} \label{eq:coord:uv}
  u = - e^{-s}, \hspace{1em}
  v = e^{-s} (e^{2x} - 1).
\end{equation}
The signs are chosen to make the arguments of the logarithm positive in
the region of interest ($v>0$, $u<0$), where the field evolution
occurs. In these coordinates the metric (\ref{eq:metric}) becomes
\begin{equation} \label{eq:metric:xs}
  g_{\mu\nu}\, dx^\mu dx^\nu =
    2 e^{2(x - s)} \left[(1 - e^{-2x}) ds^{2} - 2 ds dx\right] +
    r^2\, d\Omega^2,
\end{equation}
and the Roberts solution (\ref{eq:crit}) is simply
\begin{equation} \label{eq:crit:xs}
  r = e^{x-s}, \hspace{1em}
  \phi = x.
\end{equation}
Observe that the scalar field $\phi$ does not depend on the scale
variable $s$ at all, and the only dependence of the metric coefficients
on the scale is through the conformal factor $e^{-2s}$. This is a
direct expression of the geometric requirement (\ref{eq:ss}) in scaling
coordinates; the homothetic Killing vector $\xi$ is simply
$-\frac{\partial\ }{\partial s}$.

\section{Gauge-Invariant Perturbations of the Roberts Solution} \label{sec:gi}

Since we are ultimately concerned with the dynamics of the breaking of
the fields away from the Roberts solution, the effect due to the
growing perturbation modes, we will only consider spherically symmetric
perturbations here. Non-spherically symmetric perturbations decay
\cite{Frolov:99a} and so do not play a role in the critical behavior.
In this section, we outline how spherically symmetric perturbations of
the Roberts solution (\ref{eq:metric}) are described in gauge-invariant
formalism. A general spherically-symmetric metric perturbation is
\begin{equation}
  \delta g_{\mu\nu} dx^\mu dx^\nu =
    k_{uu}\, du^2 + 2 k_{uv}\, du\,dv + k_{vv}\, dv^2 + r^2 K\, d\Omega^2,
\end{equation}
while general perturbation of the scalar field is
\begin{equation}
  \delta\phi = \varphi.
\end{equation}
Under a (spherically-symmetric) gauge transformation generated by the
vector
\begin{equation}
  \xi^\mu = (A, B, 0, 0),
\end{equation}
the metric and scalar field perturbations transform as
\begin{equation}
  \Delta g_{\mu\nu} = \Lie_\xi g_{\mu\nu}, \hspace{1em}
  \Delta \phi = \Lie_\xi \phi.
\end{equation}
The explicit expressions for the change in the perturbation amplitudes
under the gauge transformation generated by the vector $\xi$ are
\begin{eqnarray}
  \Delta k_{uu} &=& -2 B_{,u}, \nonumber\\
  \Delta k_{uv} &=& -A_{,u} - B_{,v}, \nonumber\\
  \Delta k_{vv} &=& -2 A_{,v}, \nonumber\\
  r^2 \Delta K &=& (2u-v) A - u B, \nonumber\\
  2 r^2 \Delta \varphi &=& v A - u B.
\end{eqnarray}
Out of four metric and one matter perturbation amplitudes one can build
the total of  three gauge-invariant quantities, one describing matter
perturbations
\begin{equation}
  f = \frac{K}{2} - \varphi + \frac{1}{2u} \int k_{vv}\, dv,
\end{equation}
and the other two describing metric perturbations
\begin{equation}
  \rho = (r^2 K)_{,uv} + k_{vv} - k_{uv}
      - u k_{uu,v}/2 + (2u-v) k_{vv,u}/2,
\end{equation}
\begin{equation}
  \sigma = k_{uv} - \frac{1}{2} \int k_{vv,u}\, dv - \frac{1}{2} \int k_{uu,v}\, du.
\end{equation}
The linearized Einstein-scalar field equations
\begin{equation}
  \delta R_{\mu\nu} = 4 \phi_{(,\mu} \delta\phi_{,\nu)}, \hspace{1em}
  \delta(\Box\phi) = 0
\end{equation}
can then be rewritten completely in terms of these gauge-invariant
quantities. It is possible to show that the field equations reduce to
one master differential equation for the scalar field perturbation,
\begin{equation} \label{eq:gi:f}
  2u(u-v) f_{,uv} + (2u-v) f_{,v} - u f_{,u} - 2 f = 0,
\end{equation}
and two trivial equations relating metric perturbations to the scalar
field perturbation
\begin{equation} \label{eq:gi:sr}
  \sigma_{,u} = 2 f_{,u} + 2 f/u, \hspace{1em}
  \rho = 0.
\end{equation}
Once the gauge-invariant quantities are identified, one is free to
switch between various gauges. We conclude this section by discussing
two particularly convenient choices.

\medskip\noindent
{\bf Field gauge ($K = k_{vv} = 0$):} The scalar field perturbation
coincides with the gauge-invariant quantity $f$ in this gauge, and
expressions for other gauge-invariant quantities simplify considerably:
\begin{eqnarray}
  f &=& - \varphi, \nonumber\\
  \rho &=& - k_{uv} - u k_{uu,v}/2, \nonumber\\
  \sigma &=& k_{uv} - \frac{1}{2} \int k_{uu,v}\, du.
\end{eqnarray}
The linearized Einstein-scalar field equations are at their simplest in
this gauge, and the derivation of the master equation for $f$ above is
almost transparent. The metric and scalar field perturbation amplitudes
are trivial to obtain:
\begin{eqnarray}
  \varphi &=& - f, \nonumber\\
  k_{uv} &=& 2 f, \nonumber\\
  k_{uu,v} &=& - 4 f/u.
\end{eqnarray}

\medskip\noindent
{\bf Null gauge ($k_{uu} = k_{vv} = 0$):} This gauge was used in the
original analysis of spherically-symmetric perturbations of the Roberts
solution \cite{Frolov:97}. The motivation behind this gauge choice is
that coordinates $u$ and $v$ remain null in the perturbed spacetime.
The expressions for gauge-invariant quantities are quite simple here as
well:
\begin{eqnarray}
  f &=& K/2 - \varphi, \nonumber\\
  \rho &=& (r^2 K)_{,uv} - k_{uv}, \nonumber\\
  \sigma &=& k_{uv}.
\end{eqnarray}
For details on how to reconstruct perturbation amplitudes from
gauge-invariant quantities see \cite{Frolov:97}.

\section{Wave Propagation on the Roberts Background} \label{sec:wave}

\begin{figure}
  \epsfxsize=\textwidth \multiply\epsfxsize 4 \divide\epsfxsize 10
  \hfill\epsfbox{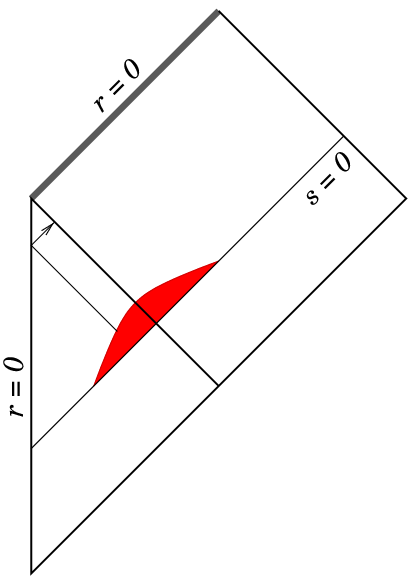}\hfill
  \medskip
  \begin{center}
    \begin{tabular}{c@{\hspace{3em}}c@{\hspace{3em}}c}
      \epsfxsize=\textwidth \multiply\epsfxsize 1 \divide\epsfxsize 5
      \epsfbox{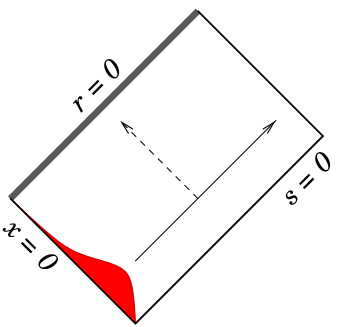} &
      \epsfxsize=\textwidth \multiply\epsfxsize 1 \divide\epsfxsize 5
      \epsfbox{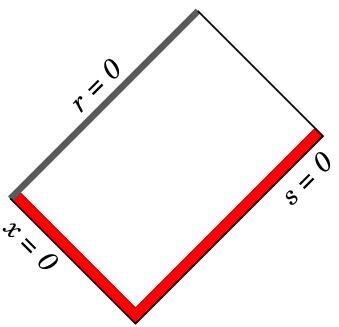} &
      \epsfxsize=\textwidth \multiply\epsfxsize 1 \divide\epsfxsize 5
      \epsfbox{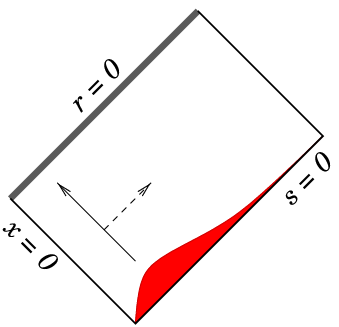}\\
      outgoing & ``constant'' & incoming
    \end{tabular}
  \end{center}
  \caption{
    Wave propagation on the Roberts background: Initial conditions can
    be equivalently specified on the surface $s=0$ extending to the
    center of the flat part of spacetime ($r=0$), or on the $(x=0) \cup
    (s=0)$ wedge. By linearity, the wavepacket can be decomposed into
    three modes: outgoing, ``constant'', and incoming.
  }
  \label{fig:wave}
\end{figure}

So, we wish to study scalar field wave propagation on the Roberts
background. Typically, one would specify initial data for the
wavepacket either on some spacelike Cauchy surface or on initial null
surface, and trace the later evolution using the field equations. Our
choice for initial surface is $u = \text{const}$ ($s=0$), which forms a
complete null surface if extended to the center of the flat spacetime
part, as illustrated in Fig.~\ref{fig:wave}. The part of a pulse
propagating through flat background evolves trivially, and can be
equivalently replaced by specifying field values on $v=0$ hypersurface.
Thus, in our problem, the initial conditions for the linearized
Einstein-scalar field equations are given on the $s=0$ surface, while
the boundary conditions are determined by junction conditions across
the null shell $v=0$, as outlined in Appendix~\ref{sec:junct}, and the
requirement that the perturbations be bounded at future infinity.

As was shown earlier, the Einstein-scalar field equations for the
spherically symmetric perturbations of the Roberts solution can be
reduced to the master differential equation
\begin{equation} \label{eq:wave}
  \D f(x,s) = 0
\end{equation}
for the single gauge-invariant quantity $f(x,s)$ describing
perturbation of the scalar field. The explicit form of differential
operator $\D$ in scaling coordinates, given by equation
(\ref{eq:gi:f}), is
\begin{equation}
  \D f=
    (1-e^{-2x})\, \frac{\partial^2 f}{\partial x^2}
    + 2\, \frac{\partial^2 f}{\partial x\, \partial s}
    + 2\, \frac{\partial f}{\partial s} - 4 f.
\end{equation}
Because of the scale-invariance of the background, the coefficients of
the differential operator $\D$ do not depend on scale, and so the
problem can be reduced to one dimension by applying a formal Laplace
transform with respect to the scale variable $s$ to all quantities and
operators. In particular, for Laplace transform of $f$ we have
\begin{equation} \label{eq:wave:L}
  F(x,k) = \int\limits_0^\infty f(x,s) e^{-ks} ds,
\end{equation}
with the inverse transformation being
\begin{equation} \label{eq:wave:L1}
  f(x,s) = \frac{1}{2\pi i}\,
    \int\limits_{\kappa-i\infty}^{\kappa+i\infty} F(x,k) e^{ks} dk.
\end{equation}
The Laplace transform can be done provided that $f$ can be bounded by
an exponential function of $s$ (that is, there exist constants $M$,
$\kappa_0$ such that $|f(x,s)| \le M e^{\kappa_0 s}$), which is a
physically reasonable condition. The contour of integration in the
complex $k$-plane for the inverse transform (\ref{eq:wave:L1}) must be
taken somewhere to the right of $\kappa_0$ ($\kappa > \kappa_0 \ge 0$).
The properties of functions of complex variables will guarantee that
the result of integration is independent of the particular contour
choice.

When applying the Laplace transform to the differential operator,
\begin{equation}
  \L_s \left[\frac{\partial f}{\partial s}\right] = k F - f(s=0),
\end{equation}
so the initial conditions of the original problem will enter as source
terms on the right hand side. Therefore the Laplace transform of the
equation (\ref{eq:wave}) is
\begin{equation} \label{eq:lwave}
  \D_k F(x,k) = h(x),
\end{equation}
where $\D_k = \L_s \D$ is now an ordinary differential operator,
algebraic in $k$, and $h$ contains information about the initial shape
of the wavepacket at $s=0$. Boundary condition on equation
(\ref{eq:lwave}) are inherited from the original problem by Laplace
transformation.

The explicit forms of the operator $\D_k$ and the relationship of $f$
to the perturbation amplitudes are the simplest when expressed in
slightly different spatial coordinate, related to the old one by
\begin{equation} \label{eq:coord:y}
  y = e^{2x} = 1 - \frac{v}{u}.
\end{equation}
The differential operator $\D_k$ is hypergeometric in nature
\begin{equation} \label{eq:wave:D}
  \D_k = y(1-y) \frac{d^2\ }{dy^2} + [1 - (k+1)y] \frac{d\ }{dy} - (k/2-1),
\end{equation}
with coefficients
\begin{eqnarray} \label{eq:coeffs}
  &c = 1,\hspace{1em}
   a+b = k,\hspace{1em}
   ab = k/2 - 1, \nonumber\\
  &a,b = 1/2\, (k \mp \sqrt{k^2-2k+4}).
\end{eqnarray}
The right hand side $h$ depends on the initial conditions as
\begin{equation} \label{eq:wave:h}
    h(y) = -y \dot{f}(y, s=0) - {\textstyle \frac{1}{2}}\, f(y, s=0).
\end{equation}
Here and later dot denotes derivative with respect to $y$ ($\dot{\ } =
d\ /dy$). Once the solution for $F$ (and hence its inverse Laplace
transform $f$) is found, one can reconstruct the other two gauge
invariant quantities $\sigma$ and $\rho$ describing the metric
perturbations using equations (\ref{eq:gi:sr}), and, in principle,
write expressions for perturbations in any desired gauge choice.

Thus, the study of wave propagation on the Roberts background is
reduced to solving the inhomogeneous hypergeometric equation
(\ref{eq:lwave}). The following analysis relies heavily on certain
properties of the hypergeometric equation, which are collected in
Appendix \ref{sec:hyper} for convenience.

So far, we have not talked about specifics of the boundary conditions
placed on the equation (\ref{eq:wave}). That depends on the physical
problem being considered. If the flat spacetime part $v<0$ were
unperturbed, the null shell junction conditions, discussed in
Appendix~\ref{sec:junct}, would require that $f=0$ on the surface
$v=0$. If some part of the pulse propagates in the flat sector $v<0$,
$f$ should be continuous across the surface $v=0$. Essentially, we can
specify value of $f$ on the wedge $(v=0) \cup (s=0)$ arbitrarily,
keeping in mind that perturbation value should be bounded at future
infinity. It is practical to split the wavepacket to three components,
as shown in Fig.~\ref{fig:wave}, and consider outgoing, ``constant''
and incoming packets separately.

\subsection{Outgoing Wavepacket} \label{sec:wave:O}

The outgoing wavepacket is characterized by
\begin{equation}
  f(x=0, s) = f_O(s), \hspace{1em} f(x, s=0) = 0
\end{equation}
and is propagating outwards to future infinity, except for backscatter
on the background curvature which goes towards the singularity at
$s=+\infty$. The boundary conditions and the initial term for equation
(\ref{eq:lwave}) are
\begin{equation}
  F(y=1, k) = F_O(k),\hspace{1em} h(y) = 0.
\end{equation}
The general solution of the homogeneous form of equation
(\ref{eq:lwave}) is
\begin{equation}
  F(y,k) = A(k) Z_1(y,k) + B(k) Z_2(y,k),
\end{equation}
where $Z_1$ and $Z_2$ are two linearly independent solutions of the
homogeneous hypergeometric equation, in notation of
Appendix~\ref{sec:hyper}. To satisfy boundary conditions at $y=1$,
parameters $A$ and $B$ must be
\begin{equation}
  A(k) = F_O(k), \hspace{1em}
  B(k) = 0.
\end{equation}
Therefore, the outgoing wavepacket solution is given by
\begin{equation}
  f(y,s) = \frac{1}{2\pi i} \int\limits_{\kappa-i\infty}^{\kappa+i\infty}
             F_O(k) \F(a,b;k;1-y) e^{ks}\, dk,
\end{equation}
where $\F$ is the hypergeometric function.

If $f_O(s)$ does not grow exponentially as $s \rightarrow +\infty$ by
itself, i.e. the image $F_O(k)$ does not have poles in $\Re\,k > 0$
half-plane, than neither does $f(y,s)$. The outgoing wavepacket just
propagates harmlessly out to future infinity, never growing enough to
cause significant deviation of the solution from the Roberts background.

\subsection{``Constant'' Wavepacket} \label{sec:wave:C}

Even more trivial is the case of the ``constant'' wavepacket,
characterized by
\begin{equation}
  f(x=0, s) = C = f(x, s=0).
\end{equation}
The boundary conditions and the initial term for equation
(\ref{eq:lwave}) are
\begin{equation}
  F(y=1, k) = C/k, \hspace{1em} h(y) = -C/2.
\end{equation}
The general solution of equation (\ref{eq:lwave}) with these boundary
conditions is
\begin{equation}
  F(y,k) = A(k) Z_1(y,k) + B(k) Z_2(y,k) + \frac{C}{k-2},
\end{equation}
and the boundary conditions at $y=1$ require that
\begin{equation}
  A(k) = - \frac{2C}{k(k-2)}, \hspace{1em}
  B(k) = 0.
\end{equation}
So the constant wavepacket solution is given by
\begin{equation}
  f(y,s) = \frac{C}{2\pi i} \int\limits_{\kappa-i\infty}^{\kappa+i\infty}
             \left[1 - \frac{2}{k}\, \F(a,b;k;1-y)\right] \frac{e^{ks}}{k-2}\, dk,
\end{equation}
and it does not grow as $s \rightarrow +\infty$ either.

\subsection{Incoming Wavepacket} \label{sec:wave:I}

By far, the most physically interesting case is the incoming wavepacket
characterized by
\begin{equation}
  f(x=0, s) = 0,\hspace{1em} f(x, s=0) = f_I(x).
\end{equation}
It propagates directly towards the singularity and is responsible for
near-critical behavior and breaking of the solution away from the
Roberts background, as we shall demonstrate. The boundary conditions
and the initial term for equation (\ref{eq:lwave}) are
\begin{equation}
  F(x=0, k) = 0,\hspace{1em} h(y) = -y \dot{f_I}(y) - f_I(y)/2.
\end{equation}
To solve the inhomogeneous hypergeometric equation (\ref{eq:lwave}),
\begin{equation} \label{eq:I}
  y(1-y) \ddot{F} + [1 - (k+1)y] \dot{F} - (k/2-1) F = h,
\end{equation}
with the boundary conditions
\begin{equation} \label{eq:I:BC}
  F(y=1, k) = 0,\hspace{1em}
  F(y=\infty, k) \text{ bounded},
\end{equation}
we must construct a Green's function out of the fundamental system of
solutions of the homogeneous equation
\begin{eqnarray} \label{eq:I:fund}
  Z_1(y) &=& \F(a,b;k;1-y) \nonumber\\
  Z_2(y) &=& (1-y)^{1-k} \F(1-a,1-b;2-k;1-y),
\end{eqnarray}
where parameters $a$ and $b$ of hypergeometric equation depend on $k$
as given by equation (\ref{eq:coeffs}). The Wronskian of the above
system is
\begin{equation} \label{eq:I:W}
  W(y) = (k-1) y^{-1} (1-y)^{-k},
\end{equation}
and the Green's function is constructed as
\begin{eqnarray} \label{eq:I:GG}
  G(y,\eta) &=& A Z_1(y) + B Z_2(y) \pm \frac{1}{2 p_0(\eta) W(\eta)}
               \left[Z_1(y) Z_2(\eta)- Z_2(y) Z_1(\eta) \right] \nonumber\\
           &=& A Z_1(y) + B Z_2(y) \pm \frac{(1-\eta)^{k-1}}{2(k-1)}
               \left[Z_1(y) Z_2(\eta)- Z_2(y) Z_1(\eta) \right],
\end{eqnarray}
where the coefficients $A$ and $B$ are to be determined by applying the
boundary conditions, and the plus-minus sign is taken depending on the
arguments of the Green's function
\begin{equation}
  \pm = \left\{
          \begin{array}{ll}
	    +, & 1 \le y \le \eta\\
	    -, & \eta \le y < \infty
	  \end{array}
	\right. .
\end{equation}
The Green's function $G$ satisfies $\D_k G(y,\eta) = \delta(y-\eta)$,
and hence can be used to construct the solution of inhomogeneous
equation
\begin{equation} \label{eq:I:F}
  F(y,k) = \int\limits_1^\infty G(y,\eta) h(\eta)\, d\eta
\end{equation}
As the initial value problem (\ref{eq:I}) is not self-adjoint, the
Green's function (\ref{eq:I:GG}) need not be symmetric in its arguments
$y$ and $\eta$. Note that the Green's function is calculated for
particular $k$-mode, and so depends on $k$, but we omitted the third
argument in $G(y,\eta;k)$ for brevity.

We now proceed to apply boundary conditions to the Green's function
(\ref{eq:I:GG}), starting with the boundary conditions at $y=1$. The
fundamental solution $Z_1$ goes to one there, while the behavior of
$Z_2$ is fundamentally different depending on the sign of $\Re (1-k)$.
If $\Re\,k < 1$, the real part of power of $(1-y)$ in (\ref{eq:I:fund})
is positive, and $Z_2$ goes to zero when $y=1$. If $\Re\,k > 1$, the
real part of power of $(1-y)$ is negative, and hence $Z_2$ diverges
when $y=1$. Substituting this into Green's function, we get
\begin{equation}
  G(y=1,\eta) = A + \frac{(1-\eta)^{k-1}}{2(k-1)}\, Z_2(\eta)
               + \left[B - \frac{(1-\eta)^{k-1}}{2(k-1)}\, Z_1(\eta) \right]
	         \left\{
		   \begin{array}{ll}
		     0, & \Re\, k < 1 \\
		     \infty, & \Re\, k > 1
		   \end{array}
		 \right\}.
\end{equation}
The boundary conditions (\ref{eq:I:BC}) require that $G(y=1,\eta)=0$,
which uniquely fixes the coefficients
\begin{equation}
  A = - \frac{(1-\eta)^{k-1}}{2(k-1)}\, Z_2(\eta), \hspace{1em}
  B = \frac{(1-\eta)^{k-1}}{2(k-1)}\, Z_1(\eta),
\end{equation}
provided $\Re\,k>1$, which is precisely the region of the complex
$k$-plane the contour of the inverse Laplace transformation should be
in. (If $\Re\,k<1$, coefficient $B$ can be arbitrary.) With these
coefficients, the Green's function (\ref{eq:I:GG}) becomes
\begin{equation} \label{eq:I:G}
  G(y,\eta) = \left\{
               \begin{array}{cl}
	         0, & y \le \eta \\
		 \frac{(1-\eta)^{k-1}}{k-1}
		   \left[Z_1(\eta) Z_2(y) - Z_2(\eta) Z_1(y)\right], &
		   \eta \le y
	       \end{array}
	     \right. .
\end{equation}
The causality of wave propagation is apparent here: the wave at $y$ is
only influenced by the data from the past $\eta \le y$.

\begin{figure}
  \centerline{
    \epsfxsize=\columnwidth \multiply\epsfxsize 4 \divide\epsfxsize 5
    \epsfbox{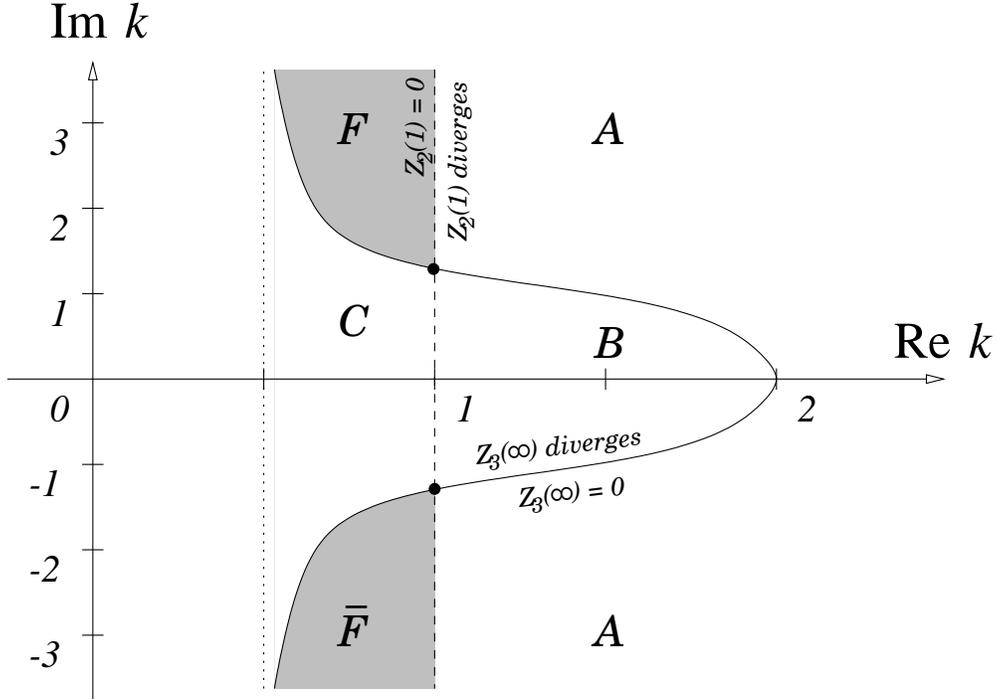}
  }
  \caption{
    Complex perturbation spectrum. Values of $k$ to the left of the solid
    line are prohibited by the boundary conditions at infinity, to the
    right of the broken line by the initial conditions at $y=1$. Values
    in the region of intersection (the shaded regions $F$ and
    $\bar{F}$) are allowed, and constitute the perturbation spectrum.
  }
  \label{fig:k}
\end{figure}

We see that the boundary conditions at $y=1$ already fix Green's
function (\ref{eq:I:G}), but we still have to satisfy the boundary
conditions at infinity! One can show that they are satisfied
automatically if (and only if) $\Re\,a \ge 0$. Curves $\Re\,k=1$ and
$\Re\,a=0$ split the complex $k$ plane into several regions, as shown
in Fig.~\ref{fig:k}. The Green's function (\ref{eq:I:G}) as written
above is defined in region $A$, but could be analytically continued to
the whole complex space. The obstructions Green's function encounters
on the boundaries between $A$ and $B$ and $A$ and $F$ are not poles,
indeed, they are not even singular for regular $(y,\eta)$. They are
rather caused by the fact that the Green's function (\ref{eq:I:G})
fails to be applicable once you cross these boundaries; in the region
$B$ boundary conditions at infinity fail to be satisfied, and in the
region $F$ free modes (solutions of homogeneous equation, that is)
exist that satisfy all boundary conditions, making the Green's function
not unique. The existence of free modes in region $F$ is at the heart
of the matter, as they grow and will determine what will happen to the
wavepacket at the later times. (It is also possible to construct
Green's function in region $C$, but since it has no bearing on our
analysis, we will not do it here.)

Once Green's function has been determined, it is simple to construct
later-time evolution of the wavepacket from the initial data using
equation (\ref{eq:I:F})
\begin{equation} \label{eq:I:soln}
  F(y,k) = \int\limits_1^y
             \frac{(1-\eta)^{k-1}}{k-1}
	     \left[Z_1(\eta) Z_2(y) - Z_2(\eta) Z_1(y)\right] h(\eta)\, d\eta.
\end{equation}
To get back from the complex $k$-plane dependence to the physical time
evolution, one performs inverse Laplace transformation
\begin{equation} \label{eq:I:soln:s}
  f(y,s) = \frac{1}{2\pi i}\, \int\limits_{\kappa-i\infty}^{\kappa+i\infty} F(y,k) e^{ks} dk.
\end{equation}
We emphasize again that a particular choice of the contour of
integration is not important, as long as it is to the right of the
obstructions on the complex plane, in our case region $F$. In practice,
one chooses the contour so that the integral (\ref{eq:I:soln:s}) is
easier to evaluate. For some approximation to work, the contour should
touch the obstruction, which means pushing it leftwards to the very
edge of region $F$ at $\Re\,k=1$.

\section{Late-time Behavior of Incoming Wavepacket} \label{sec:late}

While expressions for $f$ written down in the previous section formally
solve the problem of wave propagation on the Roberts background, they
are too complicated to be of practical use. In this section, we use the
method of the stationary phase to obtain late-time (large $s$)
asymptotic for $f(y,s)$ and analyze several physically important
regimes of the wavepacket evolution.

The method of the stationary phase deals with the approximate
evaluation of Fourier-type integrals
\begin{equation}
  f(\lambda) = \int\limits_\alpha^\beta F(k) \exp[i \lambda S(k)]\, dk
\end{equation}
for large positive parameter $\lambda$. It is based on a simple idea
that where $\exp[i \lambda S(k)]$ is oscillating extremely rapidly and
$F(k)$ is smooth, the oscillations will cancel out, and the only
contributions to the integral will be from stationary points of phase
$S(k)$, singular points of $F(k)$ and $S(k)$, and possibly end points.

Inverse Laplace transform integrals (\ref{eq:wave:L1}) are precisely of
the above type, with phase $S(k)=k$ and large parameter $\lambda$ being
the scale coordinate $s$. So the stationary phase method tells us that
the asymptotic $s \rightarrow \infty$ of the solution $f(y,s)$ is given
by singular points of $F(y,k)$ as a function of $k$, i.e. singular
points of $G(y,\eta;k)$. Therefore the study of analytic properties of
Green's function (\ref{eq:I:G}) plays a key role in understanding the
late-time evolution of the wavepacket. The possible sources of
non-analyticity in Green's function are listed below:
\begin{itemize}
  \item Branch points $k=1 \pm i\sqrt{3}$ of $a,b = 1/2 (k \pm \sqrt{k^2-2k+4})$
  \item Poles at $k=2+n$ in $Z_2$ and $k=-n$ in $Z_1$ coming from the hypergeometric function
  \item The pole at $k=1$ from the prefactor
  \item Power-law singularity of the type $(1-y)^{k-1}$
\end{itemize}
Problem with branches of the coefficients $a,b$ is absent in the
Green's function $G$ because they only enter it through the first and
second arguments of the hypergeometric function, and the hypergeometric
series are written in terms of $ab=k/2-1$ and $a+b=k$ only. Various
poles at integer values of $k$ are all canceled out because of the
antisymmetric way hypergeometric functions enter $G$. In fact, the only
source of non-analyticity in $G$ is power-law singularity, and then
only at $y,\eta \rightarrow 1$. Despite the appearance, $G(y,\eta;k)$
is an entire analytic function of $k$ provided that $y,\eta$ are
regular points.

Since the small $\eta$ region is important for the late-time evolution
of the wavepacket, it is instructive to take a closer look at the
approximation to the Green's function (\ref{eq:I:G}) there. Using the
asymptotic behavior of $Z_1$, $Z_2$ near $\eta=1$, given in
Appendix~\ref{sec:hyper}, we obtain
\begin{eqnarray} \label{eq:late:G}
  G(y, \eta\rightarrow 1)
    &\approx& \frac{(1-\eta)^{k-1}}{k-1} \left[Z_2(y) - (1-\eta)^{1-k} Z_1(y)\right] \nonumber\\
    &=& \frac{1}{k-1} \left[ \left(\frac{1-\eta}{1-y}\right)^{k-1} \F_2(y) - \F_1(y) \right],
\end{eqnarray}
where we introduce the short-hand notation
\begin{eqnarray}
  \F_1(y) &=& \F(a,b;k;1-y), \nonumber\\
  \F_2(y) &=& \F(1-a,1-b;2-k;1-y).
\end{eqnarray}
Note that even though poles in the hypergeometric functions no longer
cancel in (\ref{eq:late:G}), they are purely artifacts of the
approximation, and should be ignored.

Now we will use the above approximation (\ref{eq:late:G}) for the
Green's function to study the late-time evolution of the packet in
several important regimes.

\subsection{Evolution near $v=0$} \label{sec:late:1}

Let us first consider the behavior of the wavepacket near $y=1$, that
is, near the initial null surface $v=0$. We take the asymptotic
behavior of the initial term $h$ to be the fairly generic power law
\begin{equation} \label{eq:late:1:h}
  h(\eta) \propto (1-\eta)^\alpha,
\end{equation}
which covers the usual case of functions analytic at $y=1$ (via Taylor
expansion), as well as the case of functions with a power singularity
at $y=1$, such as free modes $Z_2$. Then the approximation to the
Green's function (\ref{eq:late:G}) gives the solution in the desired
region,
\begin{equation}
  F(y,k)
    \approx
      \int\limits_1^y \frac{1}{1-k}
        \left[ \left(\frac{1-\eta}{1-y}\right)^{k-1} \F_2(y) - \F_1(y) \right]
	\left(\frac{1-\eta}{1-y}\right)^\alpha (1-y)^\alpha\, d\eta.
\end{equation}
The above integral can be explicitly evaluated using the change of variable
\begin{equation}
  \zeta = \ln \left(\frac{1-\eta}{1-y}\right),
\end{equation}
which leads to the answer
\begin{eqnarray}
  F(y,k)
    &\approx&
      - \int\limits_{-\infty}^0 \frac{1}{1-k}
        \left[e^{(k-1)\zeta} \F_2(y) - \F_1(y)\right]
	e^{(1+\alpha)\zeta} (1-y)^{1+\alpha}\, d\zeta \\
    &=&
      - (1-y)^{1+\alpha} \frac{1}{k-1}
        \left[\frac{\F_2(y)}{k+\alpha} - \frac{\F_1(y)}{1+\alpha}\right].
\end{eqnarray}
To obtain the late-time evolution of the wavepacket near $y=1$, we use
the method of stationary phase. Observe that the only real singularity
of $F(y,k)$ is the simple pole at $k=-\alpha$, with the rest being
artifacts of the approximation (\ref{eq:late:G}). Therefore, the
late-time behavior of the wavepacket is exponential in the scale $s$;
indeed, it is given by
\begin{equation} \label{eq:late:1:f}
  f(y,s)
    \approx
      \frac{(1-y)^{1+\alpha}}{1+\alpha}\, \F_2(y;-\alpha) e^{-\alpha s}
    = \frac{1}{1+\alpha}\, Z_2(y;-\alpha)\, e^{-\alpha s}.
\end{equation}
There are several things worth noting about the above result. First,
it gives the correct answer for the evolution of the free mode. Since
$Z_2(y;p) \approx (1-y)^{1-p}$, the initial term works out to be
$h(y) \approx (1-p)(1-y)^{-p}$, so the above formula gives
$f(y,s) \approx Z_2(y;p) e^{ps}$, which is precisely what the evolution
of the free mode should be. Second, the growing modes cannot be excited
by the initial profile analytic at $y=1$. If this were the case, then
$h$ could be expanded in a Taylor series around $y=1$, each term in the
series raised to integer power corresponding to non-negative value of
$\alpha$, and so not growing at large $s$. Third, the above argument
illustrates how only the $Z_2$ content of the initial data is relevant
to the subsequent evolution of the wavepacket.

So it seems that the exponential growth of the wavepacket at late times
must be already built into the initial data in form of a power-law
divergence of the initial wave profile, just as it is encoded in the
pure free mode $Z_2$. But a power-law divergence of the perturbation
near $v=0$ causes curvature invariants to diverge at the junction,
making the surface $v=0$ a weak null singularity, which casts a shadow
of doubt on the physicality of such growing modes. The question then
arises whether it is possible to somehow eliminate the offending
divergence, while still having a wavepacket grow to large enough
values. The answer to this question is yes, and it is discussed next.

\subsection{Evolution of a wavepacket initially localized at $v=0$} \label{sec:late:local}

So what will happen if we cut off a diverging initial wave shape
(\ref{eq:late:1:h}) below some small value of $y-1$, say $\lambda$? To
put it in another way, will the power-law diverging wave localized near
$y=1$ backscatter and affect the evolution of the wavepacket at the
large $y$? If not, then subtracting such localized wavepacket from
perturbation modes discussed above will cut off the divergence at
$y=1$, while keeping the rest of the wavepacket evolution essentially
unchanged.

We model such a localized wave by adding an exponential cutoff to the
generic power law initial term (\ref{eq:late:1:h})
\begin{equation} \label{eq:late:l:h}
  h(\eta) \propto (1-\eta)^\alpha e^{(1-\eta)/\lambda}.
\end{equation}
The exponential factor is chosen because it effectively suppresses $h$
for values of $y-1>\lambda$, yet still keeps the calculations simple.
We are interested in the evolution of the wavepacket well outside the
region of initial localization, but still for small enough $y$ so that
approximation (\ref{eq:late:G}) holds, that is for $\lambda \ll y-1 \ll
1$ (that can always be arranged for small enough $\lambda$). In this
region of interest we have
\begin{equation}
  F(y,k)
    \approx
      \int\limits_1^y \frac{1}{1-k}
        \left[ \left(\frac{1-\eta}{1-y}\right)^{k-1} \F_2(y) - \F_1(y) \right]
	(1-\eta)^\alpha e^{(1-\eta)/\lambda}\, d\eta.
\end{equation}
The above integral can be evaluated by the change of variable
\begin{equation}
  t = \frac{\eta-1}{\lambda},
\end{equation}
which yields the following approximation for $F(y;k)$ in the region of
interest:
\begin{eqnarray}
  F(y,k)
    &\approx&
      \int\limits_0^\infty \frac{1}{1-k}
        \left[ \left(\frac{-\lambda t}{1-y}\right)^{k-1} \F_2(y) - \F_1(y) \right]
	(-\lambda t)^\alpha e^{-t}\, \lambda dt \\
    &=&
      - \frac{1}{k-1}
        \left[ (-\lambda)^{k+\alpha} \Gamma(k+\alpha) (1-y)^{1-k} \F_2(y)
	       - (-\lambda)^{1+\alpha} \Gamma(1+\alpha) \F_1(y) \right].
\end{eqnarray}
So outside the region of initial localization of the wavepacket, but
still for small $y$, the Laplace transform of the field perturbation is
approximately given by
\begin{equation}
  F(y,k)
    \approx
      - \frac{1}{k-1}
        \left[ (-\lambda)^{k+\alpha} \Gamma(k+\alpha) Z_2(y)
	       - (-\lambda)^{1+\alpha} \Gamma(1+\alpha) Z_1(y) \right].
\end{equation}
The main contribution to the late-time behavior is coming from the
poles of the gamma-function $\Gamma(k+\alpha)$ in the first term. Using
the stationary phase approximation, it is possible to calculate this
contribution exactly. The inverse Laplace transform of $F(y;k)$ can be
reduced to the inverse Mellin transform of the gamma-function
\begin{eqnarray}
  f(y,s)
    &\approx&
      \frac{1}{2\pi i}\, \int\limits_{\kappa-i\infty}^{\kappa+i\infty}
        \frac{Z_2(y; -\alpha)}{1+\alpha}\,
        \lambda^{k+\alpha} \Gamma(k+\alpha) e^{ks} dk\\
    &=&
      \frac{Z_2(y; -\alpha)}{1+\alpha}\, \frac{e^{-\alpha s}}{2\pi i}\,
        \int\limits_{\kappa+\alpha-i\infty}^{\kappa+\alpha+i\infty}
        \lambda^{k} \Gamma(k) e^{ks} dk\\
    &=&
      \frac{Z_2(y; -\alpha)}{1+\alpha}\, \frac{e^{-\alpha s}}{2\pi i}\,
        \int\limits_{\kappa+\alpha-i\infty}^{\kappa+\alpha+i\infty}
        \Gamma(k) \left[e^{-(s+\ln\lambda)}\right]^{-k} dk\\
    &=&
      \frac{Z_2(y; -\alpha)}{1+\alpha}\, e^{-\alpha s}\,
        [{\cal M}^{-1} \Gamma] \left(e^{-(s+\ln\lambda)}\right),
\end{eqnarray}
which is a known integral; indeed, $[{\cal M}^{-1} \Gamma](x) =
e^{-x}$. Therefore we obtain the following late-time approximation of
the field perturbation outside the region of initial localization:
\begin{equation} \label{eq:late:l:f}
  f(y,s) \approx
    \frac{\Theta(s+\ln\lambda)}{1+\alpha}\, Z_2(y;-\alpha)\, e^{-\alpha s}.
\end{equation}
This looks very similar to the earlier result (\ref{eq:late:1:f}); the
only difference is the factor $\Theta(s+\ln\lambda)$, where $\Theta$ is
defined by
\begin{equation}
  \Theta(x) = \exp\left[-e^{-x}\right],
\end{equation}
and is almost a step function, rapidly changing its value from $0$ to
$1$ as its argument becomes positive
\begin{equation}
  \Theta(x) \approx
    \left\{\begin{array}{ll}
        1, & \Re\,x>0\\ 0, & \Re\,x<0
    \end{array}\right. ,
\end{equation}
where the width of the transition is of order unity. This means that
the perturbation outside the region of initial localization does not
feel the effect of the field at $y-1<\lambda$ until a much later time,
namely $s=-\ln\lambda$, when it suddenly spreads. To put it simply, the
wavepacket initially localized at $v<\lambda$ does not backscatter
until it hits the singularity at $u=0$, and then goes out in a narrow
band $-u<\lambda$.

The above result shows how we can cut off the free mode $Z_2(y;k)$ to
avoid the curvature divergence, and yet have it grow sufficiently
large. To quantify how large can it grow, consider the free mode $f =
Z_2(y;k) e^{ks} \approx (1-y)^{1-k} e^{ks}$, with divergent scalar
curvature $R \propto f'$, and cut it off at $y-1<\lambda$. The largest
initial curvature value is of order $R \propto \lambda^{-k}$, while the
initial energy of the pulse is $\lambda R \propto \lambda^{1-k}$, and
can be made arbitrarily small. The perturbation mode will grow
exponentially until the cutoff backscatters at $s=-\ln\lambda$, at
which time its amplitude will be $\lambda^{-k}$, with proportionally
large curvatures and energies. In other words, the initial large
curvature seed localized in a small region spreads over the whole space
in the course of the evolution, with the energy of the pulse growing
correspondingly. Thus, the free perturbation modes considered here are
physical and grow exponentially to very large amplitudes, certainly
enough to leave the linear regime, and are therefore responsible for
the evolution of the solution away from the Roberts one.

\subsection{Generic initial conditions}

We now turn our attention to the evolution of the wavepacket from
generic initial conditions. It is reasonable to expect that completely
generic initial conditions will have non-zero content of all
perturbation modes present in the system, both growing and decaying, as
given by equations (\ref{eq:I:soln}) and (\ref{eq:I:soln:s}),
\begin{equation}
  f(y,s) = \int\limits_{\kappa-i\infty}^{\kappa+i\infty}
	     \left[W_2(k) Z_2(y;k) + W_1(k) Z_1(y;k)\right] e^{ks} dk.
\end{equation}
However, decaying modes will disappear very quickly, so only the growing
modes are relevant to late-time evolution. Assuming that the content of
the free growing mode $Z_2(y;k)$ is given by the weight function $W(k)$,
the generic wavepacket evolution is given by the sum of all such modes
\begin{equation}
  f(y,s) = \int\limits_\Gamma W(k) Z_2(y;k) e^{ks} dk,
\end{equation}
where the infinite contour of integration $\Gamma$ runs vertically in
the regions $F$ and $\bar{F}$ of the complex plane on Fig.~\ref{fig:k},
at $\Re\,k=1$. However, we note that the part of the contour between
endpoints $k_0^\pm = 1 \pm i\sqrt{2}$ of the regions $F$ and $\bar{F}$
does not correspond to free growing modes, as $Z_2(y\rightarrow\infty)
\approx (-y)^{-a}$ with $\Re\,a<0$ there, and so boundary conditions at
infinity are not satisfied. Therefore, for the initial wavepacket
bounded at infinity, the content of such modes is suppressed, so that
$\int dk\,[W(k)\,(-y)^{-a}] \sim 1$ for large $y$. This leads to slower
growth rates $\int dk\,[W(k)\,(-y e^s)^{-a} e^{(k+a)s}] \sim
e^{(\kappa+a)s}$ at the later times. Hence, the piece of the contour
$\Gamma$ between the endpoints $k_0^\pm$ can be omitted from the
integration without affecting the late-time evolution.

For completely generic initial conditions, we should expect $W$ to be a
smooth function of $k$ in the free mode region $F$, not preferring any
particular value of $k$. Therefore, using the stationary phase
approximation, the main contribution to the late-time behavior of the
above integral comes from the end points of the contour of integration,
\begin{equation}
  f(y,s) \approx
    - W(k_0^+) Z_2(y;k_0^+)\, \frac{e^{k_0^+ s}}{s}
    - W(k_0^-) Z_2(y;k_0^-)\, \frac{e^{k_0^- s}}{s}.
\end{equation}
Ignoring the overall weight factor, we find that the late-time evolution of
the generic wavepacket is given by
\begin{equation}
  f(y,s) \propto \Re\left[ Z_2(y;k_0)\, \frac{e^{k_0 s}}{s}\right].
\end{equation}
We emphasize that the single $k_0$-mode dominates the course of
evolution of the generic wavepacket, and thus a certain universality is
present in the way a generic perturbation departs from the Roberts
solution.

\section{Discussion} \label{sec:conclusion}

In this paper, we studied spherically symmetric perturbations of the
Roberts solution with the intent to understand how nearby solutions
depart from the Roberts one in the course of the field evolution, and
what bearing the Roberts solution has on the subject of critical
phenomena in general, and how it is related to Choptuik's critical
solution in particular. We analyzed the behavior of incoming and
outgoing wavepackets, and we focused our attention on the incoming one
as the physically relevant one for the question posed. With the aid of
the Green's function formulation, we were able to completely solve the
perturbation problem in closed form, as well as obtain simple
approximations for the late-time evolution of the field in several
important regimes.

As was shown above, the departure of the generic perturbation away from
the Roberts solution is {\em universal} in a sense that the single mode
$Z_2(y;k_0) e^{k_0 s}$ dominates the late-time evolution of the field.
The complex growth exponent gives rise to an interesting physical
effect: the perturbation developing on the scale-invariant background
evolves to have a scale-dependent structure $e^s \cos(\Im\, k_0 s)$.
The exponential growth of the amplitude of the perturbation will
eventually be stopped by the non-linear effects, while the periodic
dependence of the perturbation on the scale will most likely remain.
The period of oscillation, obtained in the linear approximation, is
\begin{equation}
  \Delta = \frac{2\pi}{\Im\, k_0}.
\end{equation}

What does this periodic dependence of the solution on the scale mean
physically? To answer this question, lets see how this symmetry is
expressed in the Schwarzschild coordinates $(r,t)$ often used in
numerical calculations (see, for example, \cite{Choptuik:93}). In this
coordinates the metric (\ref{eq:metric}) is diagonal
\begin{equation}
  ds^2 = - \alpha\, dt^2 + \beta\, dr^2 + r^2\, d\Omega^2,
\end{equation}
where metric coefficients and explicit expressions for coordinates are
given in Appendix~\ref{sec:diag}. For our purposes, it suffices to note
that the coordinate $x$ determines the ratio $r/t$, while the
coordinate $s$ sets an overall scale of both space and time coordinates
via $e^{-s}$ factor
\begin{equation}
  -\,\frac{r}{t} = \exp\left[x + \frac{1}{2}\, e^{2x}\right], \hspace{1em}
  r = \exp[x-s].
\end{equation}
One can see that taking a step $\Delta$ in the scale variable $s$ is
equivalent to scaling both spatial and time coordinates $r$ and $t$
down by a factor $e^\Delta$. Therefore, the solution being periodic in
scale coordinate $s$ is equivalent to being invariant under rescaling of
space and time coordinates $r$ and $t$ by a certain factor
\begin{equation}
  f(x, s+\Delta) = f(x, s)
  \ \Longleftrightarrow\ 
  f(e^{-\Delta} r, e^{-\Delta} t) = f(r, t).
\end{equation}
The later is an expression of the symmetry observed in the numerical
simulations of the massless scalar field collapse \cite{Choptuik:93},
and referred to as echoing, or discrete self-similarity in the literature
\cite{Choptuik:93,Gundlach:97:review}.

Thus, our simple analytical model of the critical collapse of the
massless scalar field illustrates how the continuous self-similarity of
the Roberts solution is dynamically broken to discrete self-similarity
by the growing perturbations, reproducing the essential feature of
numerical critical solutions. The value of exponent for the endpoints
of the spectrum of growing perturbation modes, $k_0 = 1 + i\sqrt{2}$,
gives the period of discrete self-similarity as
\begin{equation}
  \Delta = \sqrt{2}\pi = 4.44
\end{equation}
for linear perturbations of the Roberts solution, which is within
$25\%$ of the numerical value $\Delta=3.44$ measured by Choptuik
\cite{Choptuik:93}. Given that the perturbative model considered here
reproduces all the symmetries of the Choptuik's solution, and gives a
good estimate for the period of echoing, it is instructive to compare
the actual field profiles to the numerical calculations. There are some
technical issues connected with rewriting our results in the variables
Choptuik uses, which are addressed in Appendix~\ref{sec:diag}, but the
end result of calculation for field variable $X=\sqrt{2\pi}\,
\sqrt{\frac{r^2}{\alpha}}\, \frac{\partial\phi}{\partial r}$ from the
perturbation modes is presented in Fig.~\ref{fig:X}. Comparing this
plot to Fig.~2 in original Choptuik's paper \cite{Choptuik:93}, we see
that they share one common feature, which is the oscillatory nature of
the field solution; however, the shape of the field profiles is quite
different. This discrepancy is not surprising, however, since
perturbation methods in critical phenomena are usually viable for
calculating critical exponents, but not the field configurations
themselves.

The emerging discretely self-similar structure, and the universal way
in which the generic perturbation departs away from the Roberts
solution offer support for the conjecture that the Roberts solution is
``close'' to the Choptuik one in the phase space of all massless scalar
field configurations (in a sense of being in the basin of attraction of
the latter), and will evolve towards it when perturbed
\cite{Frolov:99a}. It seems highly unlikely, however, that the critical
mode responsible for the decay of the Choptuik solution will be
completely absent in the initial data originating near the Roberts
solution, as this usually requires fine-tuning of the parameters. So
though at first the field configuration near the Roberts solution might
seem to evolve towards the Choptuik solution, after a while the
critical mode will kick in and drive the field to either dispersal or
black hole formation. This picture is in line with the Choptuik
solution being {\it intermediate} attractor, and we offered ``ball
rolling down the stairs'' analogy earlier \cite{Frolov:99a} to
visualize the field evolution as it goes from initial configuration
(the Roberts solution) to local attractor (the Choptuik solution) and
then to global attractor (black hole or flat spacetime) in the phase
space. Unfortunately, linear perturbation methods are not sufficient to
provide a proof of the proposed scenario, and fail to give the answer
as to what would the eventual fate of the evolution be (whether black
hole or flat spacetime end-state will be selected), and how fast would
the field get there.

To completely answer these questions, one would need to employ some
sort of non-linear calculation, or perform numerical simulations of the
evolution. In particular, it would be interesting to evolve the
perturbed Roberts spacetime numerically and look for the Choptuik
spacetime as the possible intermediate attractor. Nevertheless, it may
happen that some information about Choptuik's solution can be gained
from the linear perturbation analysis of the Roberts solution. The
appeal of this method lies in the fact that such analysis could be
carried out analytically, while Choptuik's solution is still unknown in
the closed form. Similarly, one can try to study properties of other
analytically-unknown critical solutions in different matter models
based on ``nearby'' solutions with higher symmetry and simpler form.
One might also hope to obtain acceptable analytical approximations to
critical solutions, Choptuik's in particular, by going to higher order
perturbation theory in the region near the singularity.

\medskip
To summarize, our main result is the dynamical explanation of how
discretely self-similar structure forms on the continuously
self-similar background in the collapse of the minimally-coupled
massless scalar field, and the prediction for the period of this
structure $\Delta = \sqrt{2}\pi = 4.44$, which is quite close to the
numerical value $\Delta = 3.44$, considering we only did a first-order
perturbation analysis, and that non-linear effects can, in principle,
renormalize that value.

\section*{Acknowledgments}

This research was supported by the Natural Sciences and Engineering
Research Council of Canada and by the Killam Trust. I would like to
thank D. N. Page for helpful discussions of the material presented
here. Routine calculations were assisted by computer algebra engine,
{\it GRTensorII} package in particular, for which I thank the
developers.

\appendix
\section{Null Shell Junction Conditions} \label{sec:junct}

The influx of the scalar field in the Roberts solution is turned on at
the advanced time $v=0$, with the spacetime being flat before that. In
this appendix, we consider in detail junction condition across this
null surface, and their implications for boundary conditions of
perturbation problem. This discussion is adopted from general treatment
of thin null shells by Barrab\`es and Israel \cite{Barrabes&Israel:91}.

Consider the general spherically symmetric metric in null coordinates
\begin{equation}
  ds^2 = -2 e^{2 \sigma}\, du\,dv + r^2\, d\Omega^2.
\end{equation}
To fix the geometry of the soldering of spherically symmetric spacetimes
uniquely, one must match radial functions across the constant advanced
time hypersurface. To this end, we rewrite the above metric in terms of
advanced Eddington coordinates on both sides
\begin{equation}
  ds^2 = - e^\psi\, dv\, (f e^\psi\, dv - 2\, dr) + r^2\, d\Omega^2.
\end{equation}
The metric coefficients in Eddington coordinates can be calculated from
those in null coordinates by
\begin{equation}
  e^\psi = - \frac{e^{2 \sigma}}{r_{,u}}, \hspace{1em}
  f = - 2\, \frac{r_{,u} r_{,v}}{e^{2 \sigma}}.
\end{equation}
The surface density and pressure of the null shell are then determined
by jumps of the metric coefficients across the $v=\text{const}$ surface
\begin{equation}
  4\pi r^2 \epsilon = [m], \hspace{1em}
  8\pi P = [\psi_{,r}],
\end{equation}
where the local mass function $m(v,r)$ is introduced, as usual, by
\begin{equation}
  f = 1 - \frac{2m}{r}.
\end{equation}
For Minkowski spacetime $f=1$ and $\psi=\text{const}$, so below the
$v=0$ hypersurface surface we have
\begin{equation}
  m(v<0) = 0, \hspace{1em}
  \psi_{,r}(v<0) = 0.
\end{equation}
For the Roberts solution (\ref{eq:metric},\ref{eq:crit}), we have
$\sigma=0$, $r^2 = u^2 - uv$, so a direct calculation gives
\begin{equation}
  m = - \frac{uv}{4r}, \hspace{1em}
  \psi_{,r} = \frac{1}{r}\, \frac{v^2}{4r^2+v^2}.
\end{equation}
Obviously,
\begin{equation}
  \lim_{v \rightarrow +0} m = 0, \hspace{1em}
  \lim_{v \rightarrow +0} \psi_{,r} = 0,
\end{equation}
so the Roberts solution is indeed attached smoothly to the flat
spacetime, without a delta function-like stress-energy tensor
singularity associated with a massive null shell.

If one wishes to attach a perturbed Roberts spacetime to the flat one,
as we do, and still have no singularity at the junction, the null shell
matching conditions above will place boundary conditions on the
perturbation values at the junction surface. To obtain these, we take
the perturbed Roberts solution in null gauge of Ref.~\cite{Frolov:97},
given by $\sigma = \sigma(x) e^{ks}$, $r = e^{x-s} (1+\rho(x)e^{ks})$,
and calculate the null shell surface density
\begin{equation}
  [m] = [- \rho'/2 + (k/2 - 1) \rho + \sigma] e^{(k-1)s}
\end{equation}
and surface pressure
\begin{equation}
  [\psi_{,r}] = - k [ (k - 1) \rho + 2 \sigma] e^{(k+1)s}.
\end{equation}
The simultaneous vanishing of these two for arbitrary $k$ can only be
accomplished if
\begin{equation}
  \rho = \rho' = \sigma = 0,
\end{equation}
and these are precisely the boundary conditions we imposed on metric
coefficients earlier. One also has to require continuity of the scalar
field across junction, so the boundary condition on scalar field
perturbation is
\begin{equation}
  \phi = 0.
\end{equation}
These boundary conditions on perturbations in the null gauge simply
mean vanishing of gauge-invariant perturbation amplitudes (defined in
the next appendix) on the $v=0$ hypersurface.

\section{Properties of the Hypergeometric Equation} \label{sec:hyper}

We showed above that the linear perturbation analysis of the Roberts
solution can be reduced to the study of solutions of the hypergeometric
equation with certain parameters. The hypergeometric equation has been
extensively studied; for a complete description of its main properties
see, for example, \cite{Bateman&Erdelyi}. In this appendix we collect
the facts about the hypergeometric equations that are of immediate use
to us, mainly to establish notation.

The hypergeometric equation is a second order linear ordinary
differential equation,
\begin{equation}
  y(1-y) \ddot{Z} + [c - (a+b+1)y] \dot{Z} - ab Z = 0,
\end{equation}
with parameters $a$, $b$, and $c$ being arbitrary complex numbers. It has 
three singular points at $y=0,1,\infty$. Its general solution is a
linear combination of any two different solutions from the set
\begin{eqnarray}
  Z_1 &=& \F(a, b; a+b+1-c; 1-y),\nonumber\\
  Z_2 &=& (1-y)^{c-a-b} \F(c-a, c-b; c+1-a-b; 1-y),\nonumber\\
  Z_3 &=& (-y)^{-a} \F(a, a+1-c; a+1-b; y^{-1}),\nonumber\\
  Z_4 &=& (-y)^{-b} \F(b+1-c, b; b+1-a; y^{-1}),\nonumber\\
  Z_5 &=& \F(a, b; c; y),\nonumber\\
  Z_6 &=& y^{1-c} \F(a+1-c, b+1-c; 2-c; y),
\end{eqnarray}
where $\F(a, b; c; y)$ is the hypergeometric function, defined by the
power series
\begin{equation}
  \F(a, b; c; y) = \sum\limits_{n=0}^{\infty}
                   \frac{(a)_n (b)_n}{(c)_n}\, \frac{y^n}{n!},
\end{equation}
and we used shorthand notation $(a)_n = \Gamma(a+n)/\Gamma(a)$.
The hypergeometric series is regular at $y=0$, its value there is
$\F(a, b; c; 0) = 1$, and it is absolutely convergent for $|y|<1$.
Considering the hypergeometric series as a function of its parameters,
one can show that $\F(a, b; c; y_0)/\Gamma(c)$ is entire analytical
function of $a$, $b$, and $c$, provided that $|y_0|<1$.

The solutions $Z_1, \ldots, Z_6$ are based around different singular
points of the hypergeometric equation, with asymptotics given by
\begin{eqnarray}
  Z_1 = 1,\ Z_2 = (1-y)^{c-a-b} &\text{ near }& y=1, \nonumber\\
  Z_3 = (-y)^{-a},\ Z_4 = (-y)^{-b} &\text{ near }& y=\infty, \nonumber\\
  Z_5 = 1,\ Z_6 = y^{1-c} &\text{ near }& y=0.
\end{eqnarray}

Any three of the functions $Z_1, \ldots, Z_6$ are linearly dependent
with constant coefficients. In particular,
\begin{equation}
  \left[\begin{array}{c} Z_1 \\ Z_2 \end{array}\right] =
    \left[\begin{array}{cc} c_{13} & c_{14} \\ c_{23} & c_{24} \end{array}\right]
    \left[\begin{array}{c} Z_3 \\ Z_4 \end{array}\right],
\end{equation}
where the coefficient matrix is given by
\begin{equation}
  \left[\begin{array}{cc} c_{13} & c_{14} \\ c_{23} & c_{24} \end{array}\right] =
    \left[\begin{array}{cc}
      \frac{\Gamma(a+b+1-c) \Gamma(b-a)}{\Gamma(b+1-c) \Gamma(b)}\, e^{-i\pi a} &
      \frac{\Gamma(a+b+1-c) \Gamma(a-b)}{\Gamma(a+1-c) \Gamma(a)}\, e^{-i\pi b} \\
      \frac{\Gamma(c+1-a-b) \Gamma(b-a)}{\Gamma(1-a) \Gamma(c-a)}\, e^{-i\pi(c-b)} &
      \frac{\Gamma(c+1-a-b) \Gamma(a-b)}{\Gamma(1-b) \Gamma(c-b)}\, e^{-i\pi(c-a)} \\
    \end{array}\right].
\end{equation}
These relationships are true for all values of the parameters for which
the gamma-function terms in the numerators are finite, and all values
of $y$ for which corresponding series converge, with $\Im\,y > 0$. If
$\Im\,y < 0$, signs of arguments in the exponential multipliers should
be inverted. We shall not give the rest of similar relationships here.

\section{Spacetime in Curvature Coordinates} \label{sec:diag}

For comparison of our results to Choptuik's numerical simulations
\cite{Choptuik:93}, we must rewrite them in the diagonal
Schwarzschild-like coordinates Choptuik uses
\begin{equation}
  ds^2 = - \alpha\, dt^2 + \beta\, dr^2 + r^2\, d\Omega^2.
\end{equation}
One can straightforwardly check that the coordinate change
\begin{equation}
  t = - \exp\left[-s - \frac{1}{2}\, e^{2x}\right], \hspace{1em}
  r = \exp[x-s]
\end{equation}
diagonalizes the Roberts metric (\ref{eq:metric:xs}). By
self-similarity, the quantity $t/r$, as well as the metric coefficients
$\alpha$ and $\beta$ do not depend on the scale $s$, but only on the
coordinate $x$. The metric coefficients, written as functions of $x$,
are
\begin{equation}
  \alpha = 2\, \frac{\exp[e^{2x}]}{1+e^{2x}}, \hspace{1em}
  \beta = 2\, \frac{1}{1+e^{-2x}}.
\end{equation}
If one wishes, one can rewrite them as explicit functions of $t/r$,
using
\begin{equation}
  x = \frac{1}{2}\, \ln W(r^2/t^2),
\end{equation}
in terms of Lambert's $W$-function, which is defined by the
solution of transcendental equation
\begin{equation}
  W \exp(W) = x.
\end{equation}
The expressions for metric coefficients are then
\begin{equation}
  \alpha = 2\, \frac{\exp[W(r^2/t^2)]}{1+W(r^2/t^2)}, \hspace{1em}
  \beta = 2\, \frac{W(r^2/t^2)}{1+W(r^2/t^2)}.
\end{equation}
However, the coefficients $\alpha$ and $\beta$ cannot be written in
closed form in terms of elementary functions of $t/r$.

As you can see from expressions for the metric above, diagonal
Schwarzschild coordinates are not particularly well-suited for
description of the Roberts spacetime. On top of the complicated metric
form, one artifact of the diagonal coordinate system is that the null
singularity at $u=0$ gets compressed into a point at $r=t=0$. Also,
slices $t = \text{const}$ cut across the $v=0$ hypersurface, so one has
to be careful with discontinuities of the solution there.

The perturbation amplitudes in the gauge preserving diagonal form of
the metric are also quite complicated. The simplest way to get them
from gauge-invariant quantities is to explicitly a find gauge
transformation
\begin{equation}
  \xi^\mu = \left(A, B, 0, 0\right)
\end{equation}
connecting the simple field gauge $K=k_{vv}=0$ with the diagonal gauge,
fixed by conditions $K=0$ and $(2u-v)^2 k_{vv} = u^2 k_{uu}$. The
effects of the gauge transformation on the perturbation amplitudes were
given above in Section~\ref{sec:gi}. Imposing the condition $K=0$, one
finds that $B$ must be related to $A$ by
\begin{equation}
  B = \frac{2u-v}{u}\, A.
\end{equation}
$A$ is then found by imposing the other condition fixing diagonal
gauge, which leads to the following equation
\begin{equation}
  (2u-v)^2 A_{,v} - u(2u-v) A_{,u} - v A = 2u \int f\, dv.
\end{equation}
Rewriting A in scaling coordinates,
\begin{equation}
  A(y,s) = {\cal A}(y) e^{(k-1)s},
\end{equation}
transforms the above equation into the ordinary differential equation
\begin{equation}
  (1+y) \dot{\cal A} + [1 - k/2\, (1+y)] {\cal A} = - \int F\, dy,
\end{equation}
which can be easily solved to give
\begin{equation}
  {\cal A}(y) = - \frac{e^{\frac{k}{2}\, y}}{1+y}\,
                  \int\limits_1^y d\xi\, e^{-\frac{k}{2}\, \xi}
		  \int\limits_1^\xi F(\zeta)\, d\zeta.		  
\end{equation}
Once the connecting gauge transformation is known, it is trivial to
obtain the perturbation amplitudes in the Schwarzschild diagonal gauge.
In particular, the scalar field perturbation is given by
\begin{equation}
  \varphi(y; k) = - F(y; k) + {\cal A}(y; k).
\end{equation}
We end this section by observing that while the gauge transformation
term $A$ is a small correction to gauge-invariant quantities near
$y=1$, it is not at all well behaved at infinity. Indeed, it blows up
exponentially as $e^{\frac{k}{2}\, y}$! The presence of this gauge
artifact in the quite sensibly-looking diagonal gauge illustrates just
how easily one can get into trouble if one is not working in a
gauge-invariant formalism.




\begin{figure}
  \begin{center}
    \begin{tabular}{c@{\hspace{1cm}}c}
      $Z_1$ & $Z_2$ \\
      \epsfxsize=\columnwidth \multiply\epsfxsize 2 \divide\epsfxsize 5
      \epsfbox{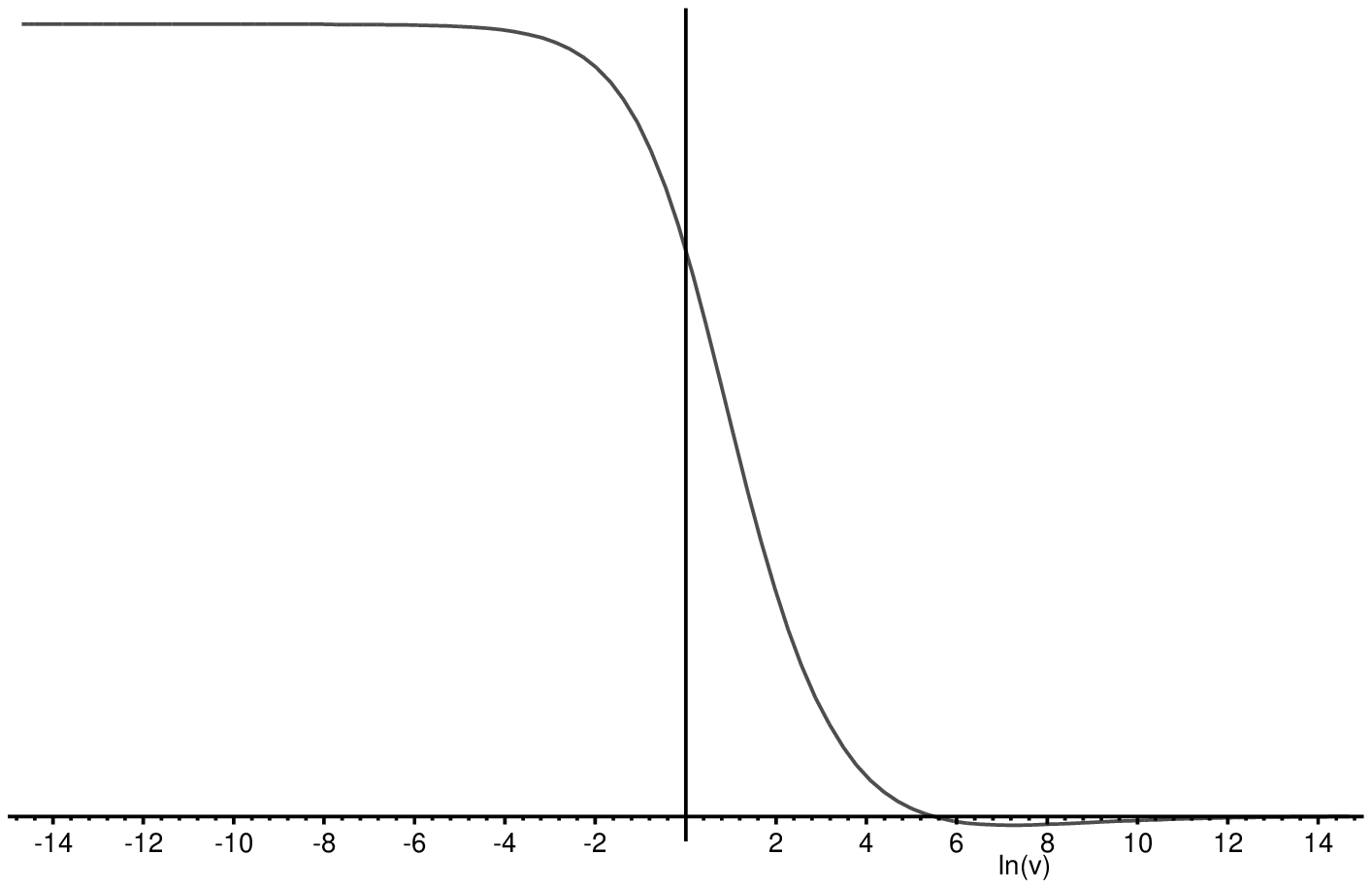} &
      \epsfxsize=\columnwidth \multiply\epsfxsize 2 \divide\epsfxsize 5
      \epsfbox{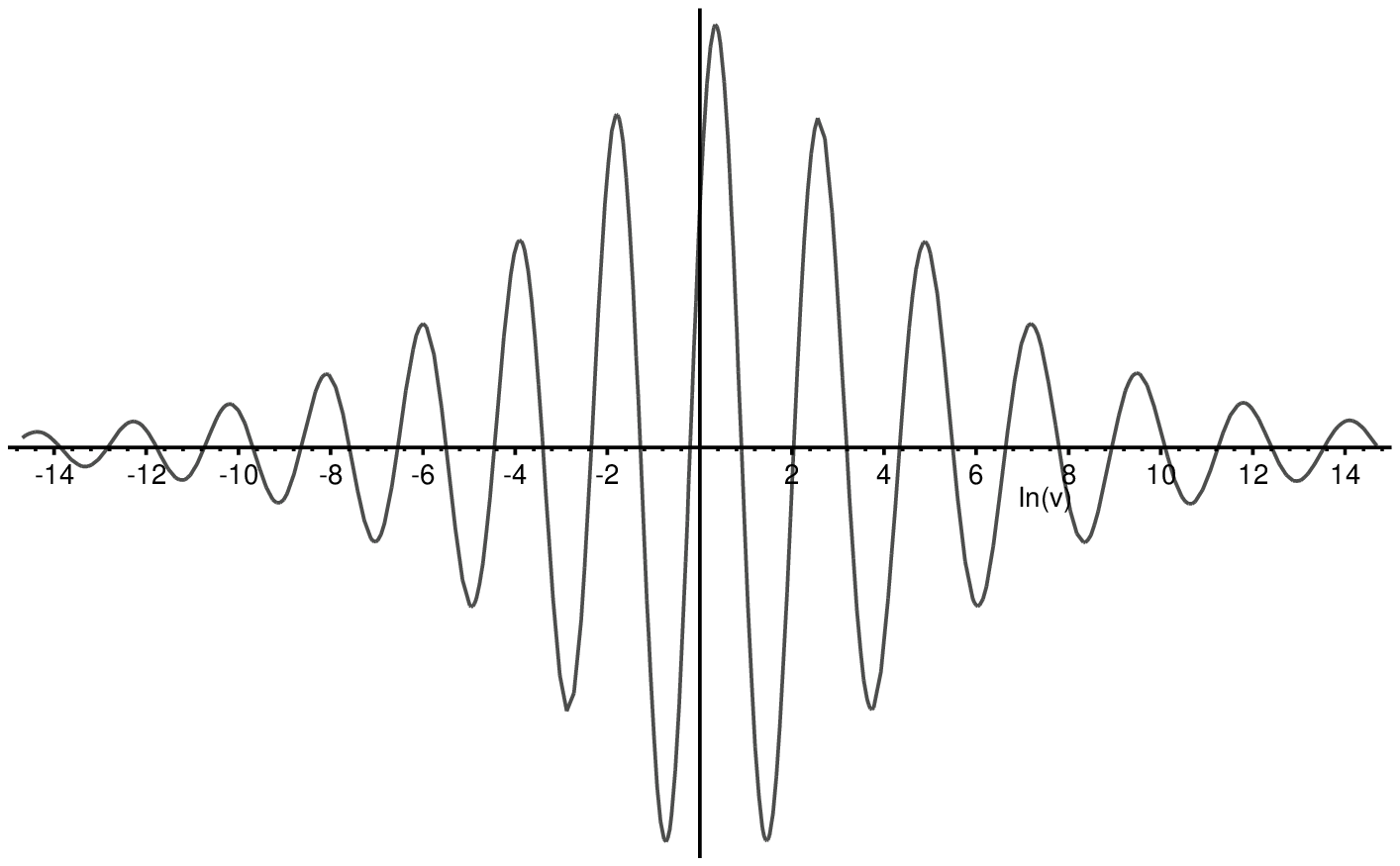} \\
    \end{tabular}\\
    (a) $k = \frac{3}{4} + 3i$\\
    \medskip
    
    \begin{tabular}{c@{\hspace{1cm}}c}
      $Z_1$ & $Z_2$ \\
      \epsfxsize=\columnwidth \multiply\epsfxsize 2 \divide\epsfxsize 5
      \epsfbox{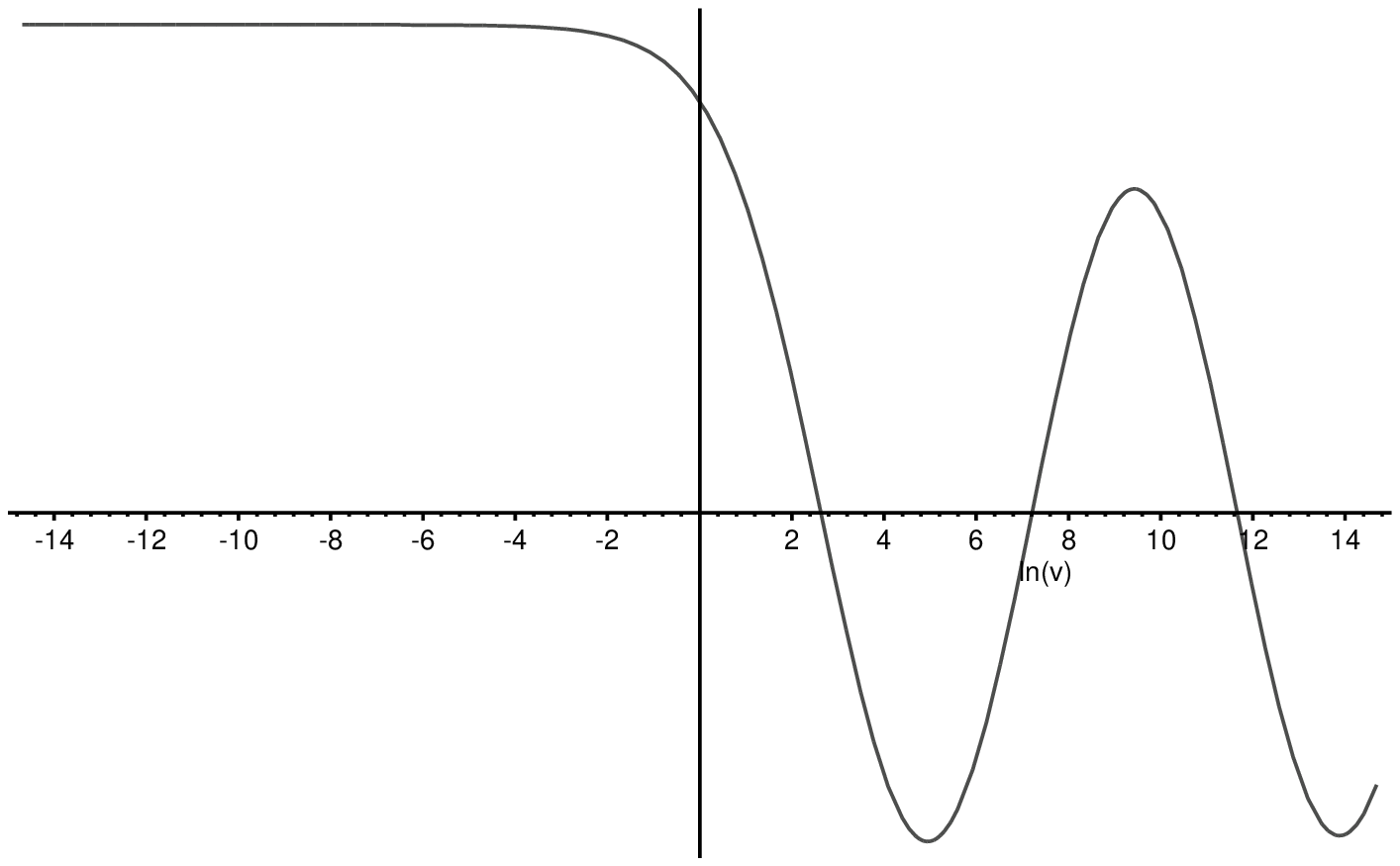} &
      \epsfxsize=\columnwidth \multiply\epsfxsize 2 \divide\epsfxsize 5
      \epsfbox{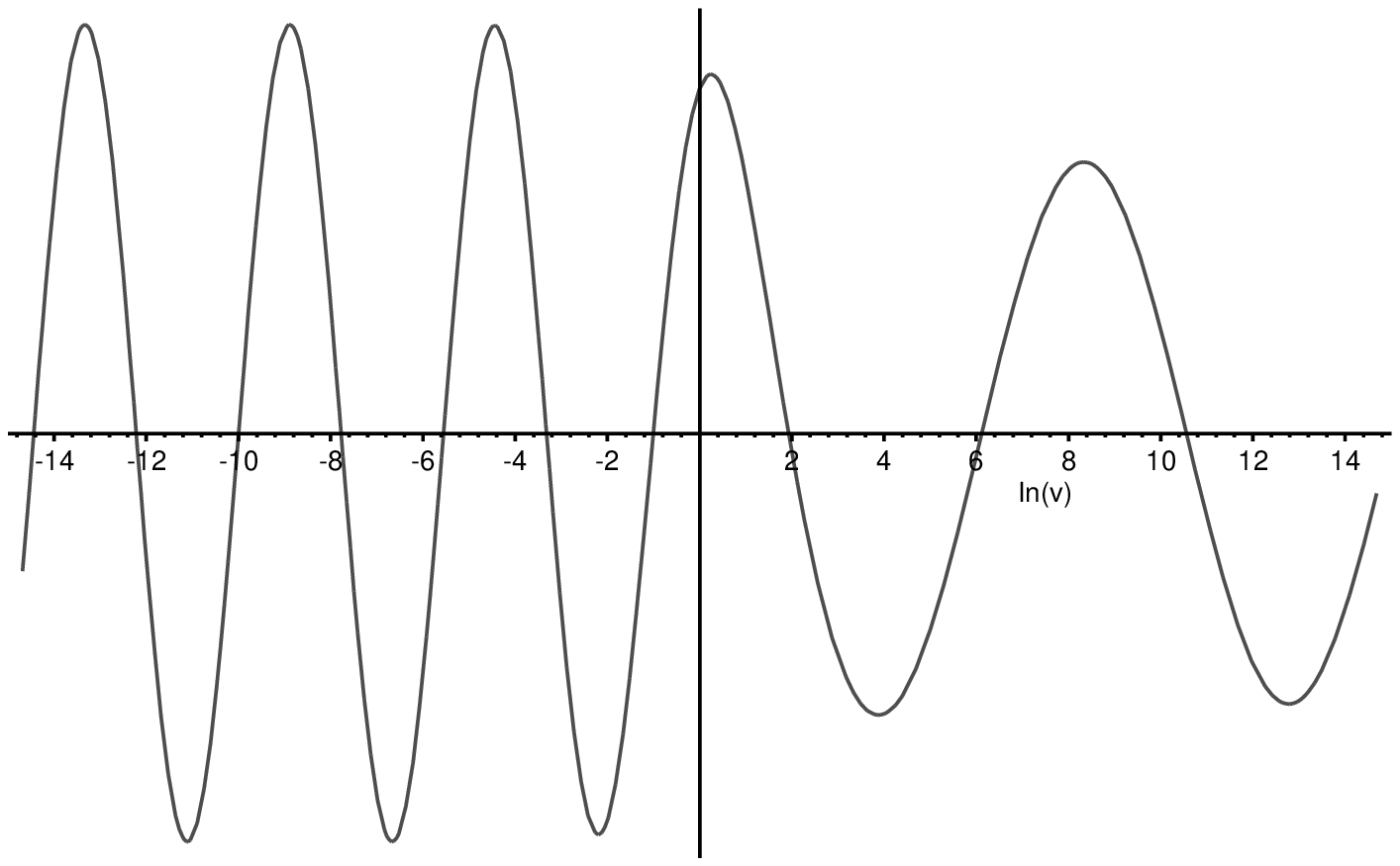} \\
    \end{tabular}\\
    (b) $k = 1 + i\sqrt{2}$\\
  \end{center}
  \caption{
    Field perturbation profiles on a slice $u=\text{const}$:
      (a) for a typical value of $k$ inside region $F$,
      (b) for a value of $k$ at the endpoint of region $F$.
    The horizontal coordinate on the plots is $\ln v$.
  }
  \label{fig:Z}
\end{figure}

\begin{figure}
  \centerline{
    \epsfxsize=\columnwidth \multiply\epsfxsize 2 \divide\epsfxsize 5
    \epsfbox{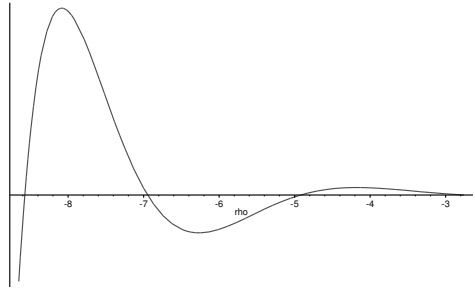}
  }
  \caption{
    Profile for the field variable $X=\sqrt{2\pi}\, \sqrt{\frac{r^2}{\alpha}}\,
    \frac{\partial\phi}{\partial r}$ on the slice $t = \text{const}$ for the
    dominant mode $f(y,s) = Z_2(y; k_0)\, e^{ks}/s$. Compare this plot
    to Fig.~2 of Choptuik's paper \protect\cite{Choptuik:93}.
  }
  \label{fig:X}
\end{figure}

\end{document}